\documentclass[11pt]{article}
\usepackage{amsmath,amssymb}
\usepackage{graphicx}
\usepackage{cite,fancyhdr}
\pdfoutput=1

\usepackage[affil-it]{authblk}

\usepackage[left=2cm,right=2cm,top=2cm,bottom=2cm,a4paper]{geometry}

\usepackage{hyperref}

\hypersetup{
pdfstartview=FitV,
colorlinks=true,
linkcolor=blue, 
citecolor=red, 
filecolor=black, 
urlcolor=blue}

\def\a{\alpha}
\def\b{\beta}

\def\L{\Lambda}

\newcommand{\KEV}{ \, {\rm keV} }

\newcommand{\GEV}{\, {\rm GeV} }
\newcommand{\TEV}{\,  {\rm TeV} }

\newcommand{\vsp}{\vspace{10pt}}

\allowdisplaybreaks

\begin{document}

\setcounter{footnote}{0}
\setcounter{figure}{0}
\setcounter{table}{0}

\title{\bf \Large Seminatural Gauge Mediation from Product Group Unification}

\author[1]{{\normalsize Hajime Fukuda}}
\author[1,2,3]{{\normalsize Hitoshi Murayama}}
\author[1]{{\normalsize Tsutomu. T. Yanagida}}
\author[4]{{\normalsize Norimi Yokozaki}}

\affil[1]{\small Kavli Institute for the Physics and Mathematics of the
  Universe (WPI), Todai Institutes for Advanced Study, University of Tokyo,
  Kashiwa 277-8583, Japan}
\affil[2]{Department of Physics, University of California,
  Berkeley, California 94720, USA} 
\affil[3]{Theoretical Physics Group, Lawrence Berkeley National
  Laboratory, Berkeley, California 94720, USA} 
\affil[4]{\small Istituto Nazionale di Fisica Nucleare, Sezione di Roma, \authorcr {\it Piazzale Aldo Moro 2, I-00185 Rome, Italy}
}

\date{}

\maketitle

\thispagestyle{fancy}
\rhead{IPMU15-0127}
\cfoot{\thepage}
\renewcommand{\headrulewidth}{0pt}

\begin{abstract}
\noindent
We propose a focus point gauge mediation model based on the product group unification (PGU),  which solves the doublet-triplet splitting problem of the Higgs multiplets. In the focus point gauge mediation, the electroweak symmetry breaking scale can be naturally explained even for multi-TeV stops.
It is known that the focus point behavior appears 
if a ratio of the number of $SU(2)$ doublet messengers to that of $SU(3)$ triplet messengers is close to $5/2$.
Importantly, this ratio (effectively) appears in our scenario based on the PGU,
if the messenger field is an adjoint representation of $SU(5)$ gauge group. 
Therefore, our focus point scenario is very predictive.
It is also pointed out the gravitino can be dark matter without spoiling the success of the thermal leptogenesis.
The absence of the SUSY CP-problem is guaranteed in the case that the Higgs $B$-term vanishes at the messenger scale.
\end{abstract}

\clearpage

\section{Introduction}
Gauge mediated supersymmetry (SUSY) breaking (GMSB)~\cite{gmsb}\,\footnote{
For early attempts, see also Refs.~\cite{gmsb_old}.
} is very attractive, since it is
free from the flavor-changing neutral current %(FCNC) 
 problem in the SUSY
standard model, which is a serious obstacle for the low-energy SUSY. 
More interestingly, most of the physical observables are predicted within the framework of renormalizable field theories, 
once a model of messenger multiplets are specified.
In fact, the spectrum of SUSY particles is unambiguously calculated using the number
of messengers and their SUSY invariant masses and SUSY breaking $B$-terms 
in a minimal GMSB.

Furthermore, it has been pointed out that 
a focus point behavior~\cite{focus_org}  
is realized in the minimal GMSB
if the numbers of $SU(2)_L$ doublet messengers $N_D$ and $SU(3)_c$ triplet messengers $N_T$
have certain values~\cite{fp_gmsb, fp_gmsb1, fp_gmsb2}~\footnote{
More precisely, in Ref.~\cite{fp_gmsb2} the focus point SUSY is achieved by fixing a combination of parameters in the superpotential of the messenger sector. Importantly, this combination is RGE invariant. Therefore, the focus point behavior is expected to be robust.
}
(see also \cite{Agashe:1999ct} for an earlier work).
It turns out that the ratio of $N_D$ to $N_T$ is always close to 5/2.
The focus point enables us to explain the origin of the electroweak symmetry breaking (EWSB) scale even when the masses of the SUSY particles are multi-TeV. Therefore, it is one of the important directions for the SUSY, after the discovery of the Higgs boson with a mass around 125 GeV~\cite{lhc_higgs}. In fact, the explanation of the observed Higgs boson mass requires large radiative corrections~\cite{higgs_susy}: it is indicated that stops are at least heavier than 3-4 TeV~\cite{higgs_3loop}, if their trilinear coupling has a moderate value.

In this paper, we show that the required ratio ($N_D/N_T \sim 5/2$) for the focus point SUSY in the minimal GMSB 
is indeed fixed by the product group
unification~\cite{pgu,pgu2}. In the product group unification, the doublet-triplet splitting problem and the rapid proton decay problem, which are very severe in the minimal $SU(5)$ grand unified theory (GUT), are easily solved.  
We show that the messenger numbers are fixed to be $(N_T, N_D)=(2,5)$ and 
the widely known fine-tuning measure $\Delta$~\cite{ft_measure} is small as 
$\Delta$=70-130 for the observed Higgs boson mass $m_h \simeq$ 125 GeV.
Moreover, we point out the gravitino can be a dark matter with a high reheating temperature while avoiding the over-closure of the universe. Thanks to this high reheating temperature, a sufficient baryon number is easily produced via the leptogenesis~\cite{leptogenesis}.
%A drawback of the model  is that all SUSY particles except for the higgsino are predicted to {\bf be heavy as $m_{\rm gluino}\gtrsim 3.5$ TeV, $m_{\rm squarks} \gtrsim 3.5$ TeV}, 
%and it is may be challenging to test this model at the LHC. 

\section{Focus point gauge mediation from product group unification}

In the focus point SUSY, the fine-tuning of the EWSB scale is significantly milder than the naive expectation  
due to a special relation among soft SUSY breaking masses: even if the mass scale of the SUSY particles 
is much larger than the EWSB scale $v_{\rm obs} \simeq 174.1$ GeV, 
the radiatively generated soft mass for the up-type Higgs is naturally close to the observed EWSB scale.
Among the focus point SUSY scenarios, the present focus point GMSB is especially attractive, since the focus point behavior is  controlled only by the number of messenger particles.

\subsection{Focus point gauge mediation} 
In the focus point gauge mediation, the required relation among the SUSY breaking masses to relax the fine-tuning is  obtained for $N_D/N_T \sim 5/2$, where $N_D$ and $N_T$ are the numbers of the $SU(2)_L$ doublet messengers and $SU(3)_c$ triplet messengers, respectively.
To see this, let us show the EWSB conditions and 
how the relevant mass parameters in the Higgs potential are written in terms of the soft mass parameters at the high energy scale. 
The EWSB scale and the ratio of the Higgs VEVs ($\tan\beta \equiv v_u/v_d$) are determined by the stationary conditions of the Higgs potential:
\begin{eqnarray}
\left[\frac{(3/5)g_1^2 + g_2^2}{4}\right] v^2 &\simeq& -\mu^2 - \frac{(m_{H_u}^2 + \frac{1}{2v_u}\frac{\partial \Delta V}{\partial  v_u}) \tan^2\beta - (m_{H_d}^2 + \frac{1}{2v_d}\frac{\partial  \Delta V}{\partial  v_d})}{\tan^2\beta-1} \Bigr|_{M_{IR}} , \nonumber \\
B_\mu (\tan\beta + \cot \beta) &\simeq& 
m_{H_d}^2 + \frac{1}{2v_d}\frac{\partial  \Delta V}{\partial  v_d}
+ m_{H_u}^2 + \frac{1}{2v_u}\frac{\partial  \Delta V}{\partial  v_u} + 2\mu^2 \Bigr|_{M_{IR}},
\end{eqnarray}
where $M_{IR}$ is taken to be the stop mass scale, $M_{IR} = (m_{Q_3} m_{\bar U_3})^{1/2}$, $g_1$ and $g_2$ are gauge coupling constants for $U(1)_Y$ (in the $SU(5)$ GUT normalization) and $SU(2)_L$, respectively, and $\Delta V$ is a one-loop effective potential. 
The soft SUSY breaking mass for the up-type (down-type) Higgs is denoted by $m_{H_u}$ ($m_{H_d}$).
Including $\Delta V$ is important to evaluate the fine-tuning for the large stop mass, and it may even be the dominant source of the fine-tuning required to obtain the observed EWSB scale.
For large $\tan\beta \gtrsim 10$, which is preferred to enhance the Higgs boson mass, $(m_{H_u}^2 + \frac{1}{2v_u}\frac{\partial  V}{\partial  v_u})$ is important to determine the EWSB scale $v$.

With this in mind, 
the soft masses for the up-type Higgs at $M_{\rm IR}$ is written in terms of the soft SUSY breaking masses at the messenger scale $M_{\rm mess}$:
\begin{eqnarray}
m_{H_u}^2(4 {\rm TeV}) &=& 0.789 m_{H_u}^2 + 0.012 m_{H_d}^2 \nonumber \\
&-& 0.236 m_{Q}^2 - 0.140 m_{\bar U}^2 - 0.032 m_{\bar E}^2 \nonumber \\
&+& 0.031  m_{L}^2 - 0.030 m_{\bar D}^2 \nonumber \\
&+& 0. 011  M_{\tilde b}^2 + 0.139 M_{\tilde w}^2  -0.253  M_{\tilde g}^2 \nonumber \\
&-&  0.017M_{\tilde{w}}  M_{\tilde{g}} - 0.002 M_{\tilde{b}} M_{\tilde{g}} ,
\end{eqnarray}
for $M_{\rm mess}=10^9$ GeV and 
\begin{eqnarray}
m_{H_u}^2(4 {\rm TeV}) &=& 0.744  m_{H_u}^2 + 0.016  m_{H_d}^2 \nonumber \\
&-& 0.288  m_{Q}^2 - 0.157  m_{\bar U}^2 -0.044  m_{\bar E}^2 \nonumber \\
&+& 0.043  m_{L}^2 - 0.041  m_{\bar D}^2 \nonumber \\
&+& 0.011 M_{\tilde b}^2 + 0.170 M_{\tilde w}^2  -0.444  M_{\tilde g}^2 \nonumber \\
&-& 0.002  M_{\tilde b} M_{\tilde w} -0.035 M_{\tilde w} M_{\tilde g} - 0.005  M_{\tilde b} M_{\tilde g} ,
\end{eqnarray}
for $M_{\rm mess}=10^{11}$ GeV, where a coefficient smaller than $10^{-3}$ is omitted. 
Soft SUSY breaking masses for $SU(2)$ doublet squarks, $SU(2)$ singlet up squarks, and down squarks are denoted by 
$m_{Q}, m_{\bar U}$ and $m_{\bar D}$, 
$m_{L}$ and $m_{\bar E}$ are soft SUSY breaking masses for left-handed and right-handed sleptons, and $M_{\tilde b}$, $M_{\tilde w}$ and $M_{\tilde g}$ are the bino, wino and gluino mass, respectively.
Here, soft SUSY breaking mass parameters in the right hand side of the above equations are defined at $M_{\rm mess}$. We take $\tan\beta=25$, $m_{t}=173.34$ GeV and $\alpha_s(m_Z)=0.1185$.

Let us consider the minimal gauge mediation model with $N_T$ pairs of the $SU(3)_c$ triplet messengers and $N_D$ pairs of the $SU(2)_L$ doublet messengers:
\begin{eqnarray}
W= \lambda_D Z \Psi_D^a \Psi_{\bar D}^a + \lambda_T Z \Psi_T^I \Psi_{\bar T}^I, \label{eq:gmsb_fundamental}
\end{eqnarray}
where $\Psi_D^a$ is a $SU(2)_L$ doublet messenger and $\Psi_T^I$ is a $SU(3)_c$ triplet messenger, and $a=1 \dots N_D$ and $I=1 \dots N_T$.
All soft SUSY breaking masses for the MSSM particles are generated from messenger loops at the messenger scale $M_{\rm mess} \simeq \lambda_D \left<Z\right> \simeq \lambda_L \left<Z\right>$ (see Appendix \ref{sec:mgmsb} for details).
Then, $m_{H_u}^2$ can be written as
\begin{eqnarray}
m_{H_u}^2(4\,{\rm TeV})&=& [-0.253  N_T^2  - 0.011  N_D N_T - 1.073 N_T  \nonumber \\
&+& 0.058  N_D^2  + 0.380 N_D] \left(\frac{\alpha_3}{4\pi} \Lambda\right)^2,
\end{eqnarray}
for $M_{\rm mess}=10^{9}$ GeV, and
\begin{eqnarray}
m_{H_u}^2(4\, {\rm TeV}) &=& [- 0.444  N_T^2  - 0.026  N_D N_T - 1.286  N_T \nonumber \\
 &+& 0.092 N_D^2  + 0.431 N_D ] \left(\frac{\alpha_3}{4\pi} \Lambda \right)^2,
\end{eqnarray}
for $M_{\rm mess}=10^{11}$ GeV. The contribution to the soft SUSY breaking masses from the messenger loops is parametrized by $\Lambda \equiv  \left<F_Z\right>/\left<Z\right>$, where $\left<Z\right>$ and $\left<F_Z\right>$ originate from the $A$-term and $F$-term of the SUSY breaking field $Z$, respectively. 
Notice that the relevant parameter $\Lambda$ is independent of unknown Yukawa couplings 
$\lambda_T$ and $\lambda_D$~\cite{fp_gmsb1}. 
The combination ${\alpha_3}\Lambda/({4\pi}) $ corresponds to the gluino mass scale.
If the messenger numbers are  taken as $(N_T, N_D)=(2,5)$, one obtain
\begin{eqnarray}
m_{H_u}^2(4\, {\rm TeV}) &=& 0.020 \left(2 \cdot \frac{\alpha_3}{4\pi} \Lambda\right)^2 , \label{eq:focus1}
\end{eqnarray}
for $M_{\rm mess}=10^9\,$GeV and 
\begin{eqnarray}
m_{H_u}^2(4\, {\rm TeV}) &=& -0.035 \left(2 \cdot \frac{\alpha_3}{4\pi} \Lambda\right)^2 , \label{eq:focus2}
\end{eqnarray}
for $M_{\rm mess}=10^{11}\,$GeV: $m_{H_u}^2(4\, {\rm TeV})$ is much smaller than the mass scale of the colored SUSY particles. Interestingly, the required numbers of $N_D$ and $N_T$ can be obtained in the product group unified (PGU) theory, which is proposed to solve the doublet-triplet splitting problem and the proton rapid decay problem. 

\subsection{Product group unification}
We consider $SU(5) \times U(3)_H$ model, where $U(3)_H \simeq SU(3)_H \times U(1)_H$.
%\begin{eqnarray}
%\frac{1}{g_1^2} = \frac{1}{g_5^2} + \frac{\mathcal{N}}{g_{1H}^2}, \ \frac{1}{g_2^2} = \frac{1}{g_5^2}, \ \frac{1}{g_3^2} = \frac{1}{g_5^2} + \frac{1}{g_{3H}^2},
%\end{eqnarray}
In the PGU, the superpotential is given by
\begin{eqnarray}
W = \mu H^i \bar H_i + H^i \bar{B}_i^\a \bar T_\a     + \bar H_i B^i_\a T^\a,
\end{eqnarray}
where $B$ and $\bar B$ are bi-fundamental fields, transforming as $B=({\bf 5}, \bar {\bf 3})$ and $\bar B=(\bar {\bf 5}, {\bf 3})$ under $SU(5) \times SU(3)_H$ gauge group, and $T$ and $\bar T$ are $SU(3)_H$ triplets, $T=({\bf 1}, {\bf 3})$ and  $\bar T=({\bf 1}, \bar {\bf 3})$. The $SU(5) \times U(3)_H$ breaks down to $SU(3)_c \times SU(2)_L  \times U(1)_Y$ by the VEV of $B$ and $\bar B$. The relevant superpotential is given by
\begin{eqnarray}
W =  Y (\bar B^\a_i B_\a^i  - 3 v_B^2) +  B_\a^i A^\a_\b  \bar B^\b_i  
\end{eqnarray}
where $Y$ and $A$ are singlet and adjoint of $SU(3)_H$, and $v_B$ is of the order of the GUT scale.
The charge assignment which is consistent with the seesaw mechanism~\cite{seesaw} is shown in Table \ref{tab:charge}.
Here, $N$ denotes a right-handed neutrino, $\Sigma_{24}$ and $\Sigma_8'$ are a messenger superfield and its ${SU}(3)_H$ counterpart, respectively, and $Z$ is a SUSY breaking field. 
Although the $U(1)_R$ is anomalous and broken at the quantum level, 
we use it for constraining the classical Lagrangian.\footnote{
Note that the $U(1)_R$ charge assignment in Table 1 does not forbids dimension-five operators ${\bf 10}{\bf 10}{\bf 10}\bar{\bf 5}$, which can lead to dangerously large proton decay rates even when suppressed by the Planck scale \cite{Murayama:1994tc,Harnik:2004yp}. However, the dimension-five operators are easily suppressed by changing the $R$ charge assignment \cite{pgu2} without spoiling the main conclusions in this paper. 
}
The superpotential of the messenger sector will be discussed later.

%In this model, the discrete $R$-symmetry, $Z_{6R}$, is anomaly free: $Z_{6R} \, SU(5)_{\rm GUT}^2$, $Z_{6R} \, SU(3)_H^2$ and $Z_{6R}\,  U(1)_H^2$ vanish. The charge assignment of the model consistent with the seesaw mechanism is summarized in the Table~\ref{tab:charges}~\cite{Evans:2011mf}. We denote the right handed superfield as $N$.
%One can also consider $Z_{4R}$ instead of $Z_{6R}$, where $Z_{4R} \, SU(3)_H^2$ anomaly vanishes, and $Z_{4R} \, SU(5)_{\rm GUT}^2$ vanishes if there is another pair of the vector-like states (${\bf 5}$ and ${\bf \bar 5}$). 

%%%%%%%%%%%%%%
\begin{table}[htp]
\caption{The charge assignment.}
\label{tab:charges}
\begin{center}
\begin{tabular}{c | c cc|cccccccc|ccc}
 & $\bar {\bf 5}$ & ${\bf 10}$ & $N$ & $H$ & $\bar H$ & $B$ & $\bar B$ & $T$ & $\bar T$ & $Y$ & $A$  & 
 $\Sigma_{24}$ & $\Sigma_8'$ & $Z$  \\
\hline \hline
$SU(5)_{\rm GUT}$  & $\bar {\bf 5}$ &  ${\bf 10}$ & {\bf 1} &
 ${\bf 5}$ & $\bar {\bf 5}$ & ${\bf 5}$ & $\bar {\bf 5}$ & {\bf 1} & {\bf 1} & {\bf 1} & {\bf 1} &  {\bf 24} & 
${\bf 1}$ & {\bf 1} \\
$SU(3)_H$  & {\bf 1} & {\bf 1} &  {\bf 1} & ${\bf 1}$ & ${\bf 1}$ & ${\bf \bar 3}$ & ${\bf 3}$ & {\bf 3} & ${\bf \bar 3}$ & {\bf 1} &${\bf 8}$ & ${\bf 1}$  & {\bf 8} & {\bf 1} \\
$U(1)_H$  & 0 & 0 &  0 & 0 & 0 & $1$ & $-1$ & $-1$ & $1$ & 0 & 0 &  0 & 0 & 0  \\
\hline
$U(1)_{R}$  & 1/5 & 3/5 & 1 & 4/5 & 6/5 & $0$ & $0$ & 4/5 & 6/5 & 2 & 2 & -1 & 3 & 4  \\
\end{tabular}
\end{center}
\label{tab:charge}
\end{table}%

%%%%%%%%%%%%%%

With the superpotential above, the bi-fundamental fields $B$ and $\bar B$ get VEVs:
\begin{eqnarray}
\left<B_\a^i\right> = v_B \delta^i_\a, \ \left<\bar B_i^a \right> = v_B \delta_i^\a.
\end{eqnarray}
As a result $SU(5) \times U(3)_H$ breaks down to $SU(3)_c \times SU(2)_L \times U(1)_Y$.
The gauge couplings are unified approximately at the GUT scale as
\begin{eqnarray}
g_1^2/g_2^2 = \left(1 + \frac{(1/15) g_5^2}{g_{1H}^2} \right)^{-1}, \ g_3^2/g_2^2 = \left(1 + \frac{g_5^2}{g_{3H}^2} \right)^{-1}, \ g_2^2 = g_5^2
\end{eqnarray}
where $g_{1H}, g_{3H} \gg g_5$ is assumed such that $g_1^2/g_2^2 \approx 1$ and  $g_3^2/g_2^2 \approx 1$ are satisfied.
The gauge couplings of $U(1)_Y$ and $SU(3)_c$ are predicted to be slightly smaller than that of $SU(2)_L$ at the GUT scale.

\subsection{Messenger sector}
Now let us consider the messenger sector. 
Since MSSM particles get soft SUSY breaking masses through messenger loops with the SM gauge interactions, 
the messenger superfields belong to representations of $SU(5)_{\rm GUT}$. 
From a small representation, following possibilities are listed: 

\vsp
(a) ${\bf 5} + \overline {\bf 5}$ case, where ${\bf 5}= {\bf (3,1) + (1,2)}$

(b) ${\bf 10} + \overline {\bf 10}$ case, where ${\bf 10}={\bf (3,2) + (\bar{3},1) + (1,1)}$

(c) ${\bf 15} + \overline {\bf 15}$ case, where ${\bf 15}={\bf (1,3) + (\bar{3},2) + (6,1)}$

(d) ${\bf 24}$ case, where  ${\bf 24}= {\bf (1,3) + (8,1) + (\bar{3},2) + ({3}, 2) +  (1,1)}$, \vsp \\ 
where $(n_1, n_2)$ corresponds to $n_1$ and $n_2$ dimensional representations of $SU(3)_c$ and $SU(2)_L$.
Larger representations are not attractive since they lead to the Landau pole below the GUT scale, unless the messenger scale is sufficiently high.

If all of the messengers participate in generation of the soft SUSY breaking masses for the MSSM particles,
$N_D/N_T=1$ in all the cases.  However, $SU(2)_L$ singlets can be made heavy by the mechanism described below in the PGU, so that we effectively get $N_D/N_T \gg 1$. Then,
the case (a) corresponds to $(N_T, N_D)=(0,1)$.
For (b) we can have $(N_T, N_D)=(2,3)$ if the messenger in the representation ${\bf (\bar 3,1)}$ (and ${\bf (3,1)}$) is heavy. 
However, $N_D/N_T = 1.5$ is too small to realize the focus point.
For (c), one can obtain $(N_T, N_D)=(2,7)$ if the messenger in ${\bf (6,1)}$ and its vector-like partner are heavy. 
In this case, $N_D/N_T$ is too large.
Finally in the case (d),  
 we obtain the sparticle mass spectrum corresponding to $(N_T, N_D)=(2,5)$, if the octet messenger in ${\bf(8,1)}$ representation is heavy. 
Based on these considerations, we choose the messenger superfields in the adjoint representation of $SU(5)_{\rm GUT}$.
This is a clear advantage over the models in Refs.~\cite{fp_gmsb,fp_gmsb1,fp_gmsb2}, where the required numbers of $(N_T, N_D)$ are chosen by hand.

The superpotential in the messenger sector is given by
\begin{eqnarray}
W = \lambda_{24} Z (\Sigma_{24})^i_j (\Sigma_{24})^j_i + \frac{c_0}{M_P} \bar B_i^\a {(\Sigma_{24})}^i_j  {(\Sigma_8')}^\b_\a B^j_\b, \label{eq:mess_mass}
\end{eqnarray}
where $Z$ is a SUSY breaking field, and the charge assignment of the fields is shown in Table \ref{tab:charge}.
 %and $M_{24} \sim M_8'$ is assumed. 
The first term gives 
\begin{eqnarray}
 \lambda_{24} \left(\left<Z\right> + \left<F_Z\right> \theta^2 \right) \Bigl( X \bar X + {\rm Tr} (\Sigma_3^2) + {\rm Tr} (\Sigma_8^2) \Bigr),
\end{eqnarray}
where $X$, $\bar X$, $\Sigma_3$ and $\Sigma_8$ correspond to ${\bf ({3}, 2)}$, ${\bf (\bar{3},2)}$, ${\bf (1,3)}$ and ${\bf (8,1)}$, and $U(1)_Y$ charges of $X$ and $\bar X$ are -5/6 and 5/6.
From the second term in Eq.(\ref{eq:mess_mass}), the octet messenger $\Sigma_8$ has as a SUSY-invariant Dirac mass with $\Sigma_8'$, $W\ni \mu_8 {\rm Tr}(\Sigma_8 \Sigma_8')$. This dirac mass is estimated as $\mu_8 = 10^{12}$-$10^{14}$ GeV, depending on $v_{B}$ and $c_0$.
%
%\begin{eqnarray}
%W &=& M_{24} (\Sigma_{24})^i_j (\Sigma_{24})^j_i  + \lambda_{24} Z (\Sigma_{24})^i_j (\Sigma_{24})^j_i \nonumber \\
%&+& M_8' (\Sigma_{8}')^\a_\b  (\Sigma_{8}')_\a^\b + \frac{c_0}{M_P} \bar B_i^\a {(\Sigma_{24})}^i_j  {(\Sigma_8')}^\b_\a B^j_\b \nonumber \\
%&\rightarrow&  M_{24}{\rm Tr}(\Sigma_8^2) + M_8' {\rm Tr}({\Sigma_8'}^2) + \mu_{8} {\rm Tr}(\Sigma_8 \Sigma_8'),
%\end{eqnarray}
The mass eigenvalues are 
\begin{eqnarray}
M_{\pm}^2 = \mu_8^2 \left[ 1 + \frac{1}{2} k^2 \pm \frac{1}{2}k \sqrt{4+k^2} \right],
\end{eqnarray}
where $k= \lambda_{24} \left<Z\right>/\mu_8$. Provided that $\mu_8 \gg \lambda_{24} \left<Z\right>$,
the mass spectrum of the MSSM particles is determined by the light messengers, $X$, $\bar X$ and $\Sigma_3$. 
The relevant part of the messenger sector is given by
\begin{eqnarray}
W = \lambda_X \left(\left<Z\right> + \left<F_Z\right> \theta^2 \right) X \bar {X} + \lambda_3 \left(\left<Z\right> + \left<F_Z\right>\theta^2 \right) {\rm Tr} (\Sigma_3^2) ,
\end{eqnarray}
where $\lambda_X \sim \lambda_3$. 
%At the GUT scale, $\lambda_X=\lambda_3=\lambda_{24}$ and $M_3=M_X=M_{24}$ are satisfied, and hence, the relation
%\begin{eqnarray}
%\lambda_X/M_X = \lambda_3/M_3,
%\end{eqnarray}
%is hold at the any scale. 
Then, we have 
\begin{eqnarray}
M_{\tilde w} &\simeq& \frac{\a_2}{4\pi} (5 \Lambda), \ \ M_{\tilde g} \simeq \frac{\a_3}{4\pi} (2 \Lambda),  \nonumber \\
m_Q^2 &\simeq& \frac{8}{3} \left(\frac{\a_3}{4\pi}\right)^2 (2\L^2) + \frac{3}{2} \left(\frac{\a_2}{4\pi}\right)^2 (5\L^2), \nonumber \\
m_{\bar U}^2 &=& m_{\bar D}^2 \simeq \frac{8}{3} \left(\frac{\a_3}{4\pi}\right)^2 (2\Lambda^2), \nonumber \\ 
m_{H_u}^2 &=& m_{H_d}^2 = m_{L}^2 \simeq \frac{3}{2}  \left(\frac{\a_2}{4\pi}\right)^2 (5 \Lambda^2),
\end{eqnarray}
where we have neglected the contributions proportional to $\a_1$ and $\a_1^2$.
Apart from the $U(1)_Y$ contributions, the generated mass spectrum is essentially the same as that in the minimal gauge mediation model defined by Eq.(\ref{eq:gmsb_fundamental}) with $(N_T, N_D)=(2,5)$.
 %in the gauge mediation model with $N_T$ pairs of the $SU(3)_c$ triplet and $N_D$ pairs of the $SU(2)_L$ messengers.
Therefore, as shown in Eq.(\ref{eq:focus1}) and (\ref{eq:focus2}), this gauge mediation model using adjoint messengers 
significantly reduces the fine-tuning of the EWSB scale.

\begin{figure}[!t]
\begin{center}
\includegraphics[scale=1.1]{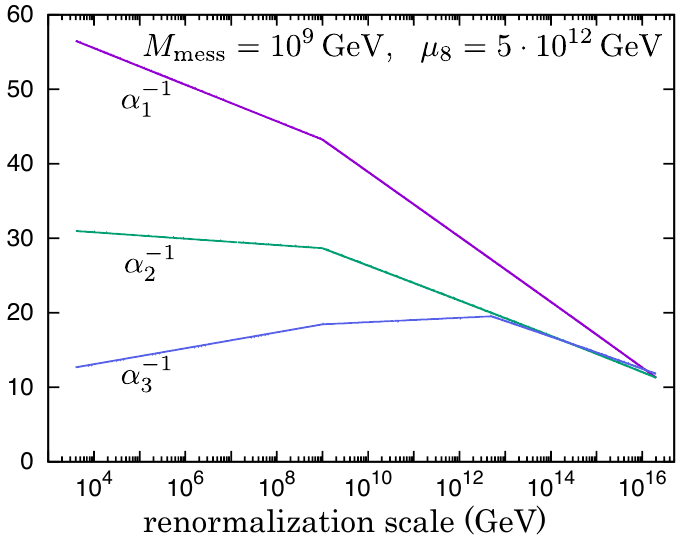}
\caption{The running of the gauge coupling with adjoint messengers.
We use two-loop renormalization group equations and take $m_{\rm SUSY}=4$\,TeV.
}
\label{fig:running}
\end{center}
\end{figure}

With the messenger multiplets $\Sigma_{24}$ and the additional multiplet $\Sigma_8'$, 
the evolution of the gauge couplings are shown in Fig.~\ref{fig:running}, neglecting small mass splitting of $M_{\pm}$, i.e. $k=0$. We use two-loop renormalization group equations~\cite{Martin:1993zk}.
The gauge coupling unification holds in a non-trivial way.
%%%%%%%%%%%%%%%
%%%%%%%%%

\section{The fine-tuning and mass spectrum}

Let us numerically estimate the fine-tuning $\Delta$ in the region 
where the observed Higgs boson mass is explained. We also
 show the mass spectra of the MSSM particles in the relevant region.
In numerical calculation, we use {\tt softsusy 3.6.1}~\cite{softsusy} to evaluate 
the SUSY mass spectra. 
The Higgs boson mass is calculated using {\tt FeynHiggs 2.11.2}~\cite{feynhiggs}.

\subsection{The fine-tuning}
To estimate the fine-tuning, we employ the following fine-tuning measure~\cite{ft_measure}:~\footnote{
The result of the derivative with respect to the messenger scale is similar to the derivative with respect to $|F_Z|$. This is because
\begin{eqnarray}
\frac{\partial }{\partial \ln M_{\rm mess}} \ln v = 
\Bigl( -\frac{\partial }{\partial \ln \Lambda} + \frac{\partial }{\partial \ln M_{\rm mess}} \Bigr|_{{\rm fixed}\, \Lambda} \Bigr) \ln v .
\end{eqnarray}
}
\begin{eqnarray}
\Delta = {\rm max} \{ \Delta_a \}, \ \ \Delta_a = \Bigl\{ \frac{\partial \ln v}{\partial \ln \mu}, \frac{\partial \ln v}{\partial \ln |F_Z|}, \frac{\partial \ln v}{\partial \ln B_{\rm mess}} \Bigr\}_{v=v_{\rm obs}},
\end{eqnarray}
where $v_{\rm obs} \simeq 174.1$ GeV, and $B_{\rm mess}$ is the Higgs $B$-term at the messenger scale, 
$B_\mu/\mu\,|_{M_{\rm mess}}$. Here, $V \ni B_\mu H_u H_d + h.c.$ and $W \ni \mu H_u H_d$.

\begin{figure}[!t]
\begin{center}
\includegraphics[scale=1.2]{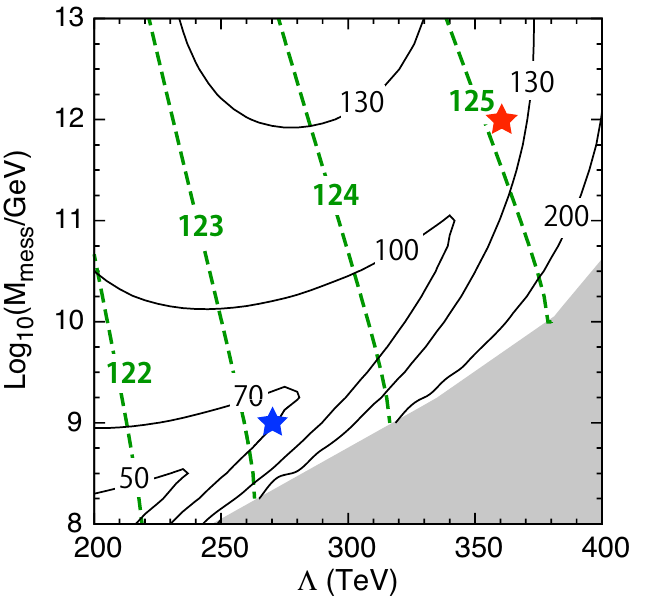}
\caption{The Higgs boson mass in the unit of GeV (green) and $\Delta$ (black) in the gauge mediation model with adjoint messengers.
We take $\tan\beta=25$, $m_t({\rm pole})=173.34$\,GeV and $\alpha_s(m_Z)=0.1185$.}
\label{fig:pgu_gmsb}
\end{center}
\end{figure}

Now, let us show 
the required $\Delta$ to explain the observed Higgs boson mass around 125 GeV is significantly reduced. 
In Fig.~\ref{fig:pgu_gmsb}, we show the contours of the calculate Higgs boson mass $m_h$ and $\Delta$. The top pole mass is taken to be $m_t ({\rm pole})=174.34$ GeV and $\alpha_s(m_Z)=0.1185$.
The observed Higgs boson mass is explained with a mild fine-tuning $\Delta=70$\,-$130$.
Here, we comment on the consistent range of $m_h$ with the measured value.
The combined measurement of the Higgs boson mass by the ATLAS and CMS 
has an uncertainty $\pm$0.5 GeV at 2$\sigma$ level~\cite{lhc_higgs}, and 
the experimental uncertainty in the top mass measurement is $\pm 1.5$\,GeV at 2$\sigma$ level~\cite{top_mass_th}.
Moreover, the theoretical uncertainty estimated by {\tt FeynHiggs  2.11.2} is about $\pm 1$\,GeV for the Higgs boson mass. 
 In total, we consider the region of $m_h \gtrsim 123$\,GeV to be consistent with the measured Higgs boson mass.
%Thus in all the plots we show the Higgs boson mass in the range 123-125 GeV.

\begin{figure}[!t]
\begin{center}
\includegraphics[scale=1.2]{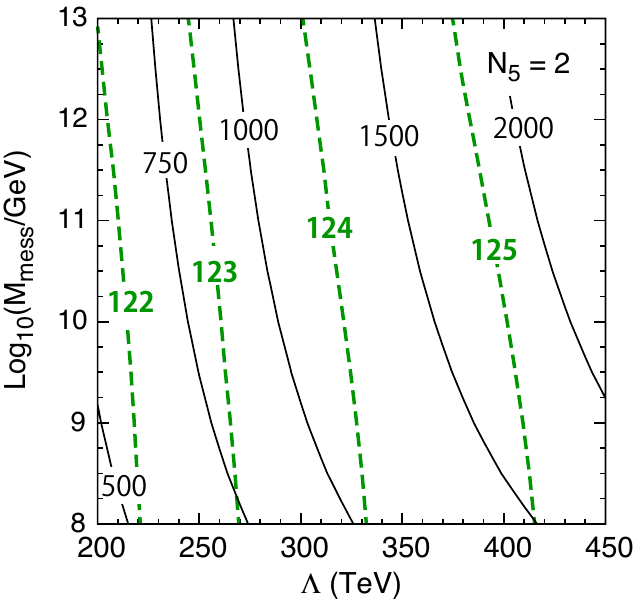}
\caption{The Higgs boson mass and $\Delta$ in the minimal gauge mediation with $N_5=2$. The other parameters are the same in Fig.~\ref{fig:pgu_gmsb}.}
\label{fig:gmsb_minimal}
\end{center}
\end{figure}

For comparison, we also show that the results in the minimal gauge mediation model with $N_5=N_T=N_D=2$ in Fig.~\ref{fig:gmsb_minimal}.  In the region $m_h=123$\,-\,125\,GeV, $\Delta=750$\,-\,$1500$ for $M_{\rm mess}=10^{\,8\mbox{-}9}$ GeV. We see that the fine-tuning in our focus point gauge mediation model 
is about ten times better than the minimal gauge mediation model with the complete $SU(5)$ multiplets.

%%%%%%%%%%%%%%%
%%%%%%%%%

%%%%%%%%%%%%%%%
%%%%%%%%%

\subsection{Mass spectra of the SUSY particles}

Here, we list $\Delta$ and mass spectra of the SUSY particles for different model points in Table. \ref{table:mass}. 
The model point {\bf P1} ({\bf P2}) is shown as the blue (red) star in Fig.~\ref{fig:pgu_gmsb}.
It can be seen that the observed Higgs boson mass is explained for $m_{\rm stop}=3$-5\,TeV. In this case, the gluino mass is as heavy as $3.5$\,-\,$5$ TeV.
Still, the fine-tuning $\Delta$ is as mild as 70\,-\,120.

%%%%%%%%%%%%%%%%%%%%%%%%%%%%%%%%%%%%%%%%%%%%%
\begin{table}[t]
    \caption{Mass spectra and $\Delta$ for different model points.
    }
      \begin{center}
    \begin{tabular}{  c | c  }
    {\bf P1} & \\
    \hline
    $M_{\rm mess}$ & $10^9$ GeV \\
    $\Lambda$ & 270 TeV \\
    $\tan\beta$ & 25 \\
    \hline
\hline    
    $m_{h}$ & 123.2\,GeV \\
    $\Delta $ & 67 \\
    $|\Delta_{\mu}| $ & 56 \\
    $\mu $ & 492 GeV \\
    $m_{3/2} $ & 64 keV/$\lambda_{24}$ \\
\hline
    $m_{\rm gluino}$ & 3.6 TeV \\
    $m_{\rm squark}  $ & 3.7\,-\,4.5 TeV \\
    $m_{\rm stop}  $ & 3.1, 4.2 TeV \\
    $m_{A}  $ & 2.5 TeV \\
    $m_{\tilde{e}_L} (m_{\tilde{\mu}_L})$ & 2.7 TeV\\
    $m_{\tilde{e}_R} (m_{\tilde{\mu}_R})$ & 1.4 TeV \\
    $m_{\tilde{\tau}_1}$ & 1.3 TeV \\
    $m_{\chi_1^0}$ & 503 GeV \\
     $m_{\chi_1^{\pm}}$ & 505 GeV \\
     $m_{\chi_2^{\pm}}$ & 3.4 TeV \\
    \end{tabular}
    %%%%%%%%%%%%%%%%%%%%%
        \begin{tabular}{  c | c  }
     {\bf P2} & \\
    \hline
        $M_{\rm mess}$ & $10^{12}$ GeV \\
    $\Lambda$ & 360 TeV \\
    $\tan\beta$ & 25 \\
    \hline
\hline    
    $m_{h}$ & 125.1\,GeV \\
    $\Delta $ & 109 \\
    $|\Delta_{\mu}| $ & 108 \\
     $\mu $ & 685 GeV \\
    $m_{3/2} $ & 86 MeV/$\lambda_{24}$ \\
\hline
    $m_{\rm gluino}$ & 4.7 TeV \\
    $m_{\rm squark}  $ & 4.7\,-\,6.1 TeV \\
    $m_{\rm stop}  $ & 3.6,  5.5 TeV \\
    $m_A$ & 3.7 TeV \\
    $m_{\tilde{e}_L} (m_{\tilde{\mu}_L})$ & 4.1 TeV\\
    $m_{\tilde{e}_R} (m_{\tilde{\mu}_R})$ & 2.4 TeV \\
    $m_{\tilde{\tau}_1}$ & 2.2 TeV \\
    $m_{\chi_1^0}$ & 701 GeV \\
     $m_{\chi_1^{\pm}}$ & 703 GeV \\
     $m_{\chi_2^{\pm}}$ & 4.6 TeV \\
    \end{tabular}
  \label{table:mass}
  \end{center}
\end{table}
%%%%%%%%%%%%%%%%%%%%%%%%%%%%%%%%%%%%%%%%%%%%%

\section{The gravitino dark matter}

Finally, the gravitino dark matter is discussed. 
We show that the gravitino can be dark matter 
without spoiling the success of the thermal leptogenesis.

In gauge mediation models, the gravitino is the lightest SUSY particle and a dark matter candidate.
The abundance of the gravitino can be estimated for different two cases: $T_R > T_f$ and $T_R < T_f$, where
$T_R$ is the reheating temperature and $T_f$ is the freeze-out temperature of the gravitino, 
\begin{eqnarray}
T_f \simeq 9.3 \TEV \left(\frac{g_*(T_f)}{230} \right)^{1/2} \left(\frac{m_{3/2}}{100 \KEV} \right)^2 \left( \frac{5\TEV}{m_{\tilde g}}\right)^2.
\end{eqnarray}

If the reheating temperature is smaller than $T_f$, 
the gravitino is produced mainly from the thermal scattering and the decay of superparticles\,\cite{Moroi:1993mb}. In this case, numerical analysis shows us that the reheating temperature should be lower than about $100$\,GeV--$1$\,TeV in order that gravitino does not overclose the Universe\,\cite{Cheung:2011nn}.
%In this case, the abundance is estimated as
%\begin{eqnarray}
%(\Omega_{3/2})_{\rm tsc} h^2 = 0.15 \cdot   \left( \frac{T_R}{50 \GEV} \right) \left( \frac{100 \KEV}{m_{3/2}} \right) \left(\frac{m_{\tilde g}}{5\TEV}\right)^2,
%\end{eqnarray}
%and the observed dark matter abundance is explained if $T_R$ is sufficiently low.

%There may be sizable amount of the gravitino is produced by the freeze-in:
%\begin{eqnarray}
%(\Omega_{3/2})_{fi} h^2 = \dots
%\end{eqnarray}

The second case, $T_R > T_f$, may be more attractive since the observed baryon number is explained by the thermal leptogenesis\cite{leptogenesis}. With this high reheating temperature, the gravitino is thermalized and the relic abundance is estimated as
\begin{eqnarray}
\Omega_{3/2}^{th} h^2 \simeq 50 \cdot \left(\frac{m_{3/2}}{100 \KEV} \right) \left(\frac{230}{g_* (T_f)} \right),
\end{eqnarray}
which does not depend on $T_R$. 
Therefore, for $m_{3/2} \simeq 100\KEV$, if we obtain a dilution factor, $D\simeq 490$, from the extra entropy production,  the observed dark matter abundance is explained. 
This dilutes the baryon asymmetry as well, but that generated by the thermal leptogenesis can be large enough as discussed later.
% and the reheating temperature $T_R$ is around that region.

The significant entropy production may occur for $T_R > M_{24} (\equiv \lambda_{24} \left<Z\right>)$\,\cite{Fujii:2002fv, Hamaguchi:2014sea}. 
Provided that the decay temperature of the messenger $T_d$ is sufficiently small,~\footnote{
	To be more precise, the messenger field dominates the energy density of the universe at the temperature~\cite{Fujii:2002fv}
	\begin{eqnarray}
	T_c \simeq \frac{4}{3} M_{\rm mess} Y_{\rm mess} \simeq 490\GEV \left(\frac{M_{\rm mess}}{10^9\GEV}\right)^2,
	\end{eqnarray}
	where $Y_\text{mess}$ is the number density of the messengers divided by the entropy density. Hence, $T_c  >  T_d$ is required.
} 
the energy density of the messenger once dominates the energy density of the universe
after the gravitino is freeze-out.
%Subsequently, it decays via the following superpotential:
Then, it decays via the superpotential:~\footnote{
There also exists
$K = (1/M_P) H \Sigma_{24} \bar {\bf 5} + h.c.$ and $K = (1/M_P) \bar H \Sigma_{24} (\bar {\bf 5}_i)^\dag + h.c.$.
However, the former becomes $W \sim (m_{3/2}/M_P) H \Sigma_{24} \bar 5$ after the K\"ahler transformation, 
which leads to a very small decay width and, because of the equation of the motion, the latter is equivalent to Eq.\,(\ref{eq:messdecay}) with the down-type Yukawa suppression.
}
\begin{eqnarray}
\label{eq:messdecay}
W_\text{decay} = \frac{k_0}{M_P} \bar H^2 {\bf 10} \Sigma_{24}.
\end{eqnarray}
The decay of the octet messenger $\Sigma_8$ is suppressed by the colored Higgs mass, and, we thus assume it is heavier than the reheating temperature and not created, so that we hereafter consider 
the decays of the ${SU}(2)$ charged messengers, $X$, $\bar X$ and $\Sigma_3$. 
In the following, we calculate the scalar messenger decay. Note that we have to take into account not only the interaction with fermions arising sorely from $W_\text{decay}$, but also the $F$-term potential involving the effective mass term of $\Sigma_{24}$. The total decay width is
\begin{eqnarray}
\label{decaywidth24}
\Gamma_\text{mess} \simeq 10^{-4}{k_0}^2 \frac{{M_\text{mess}}^3}{{M_\text{P}}^2}.
\end{eqnarray}
The dilution factor is estimated as
\begin{eqnarray}
D \left(\equiv \frac{s_{\rm after}}{s_{\rm before}}\right) \simeq \frac{4}{3} \frac{M_{\rm mess} Y_{\rm mess}}{T_d} \simeq  490 \left(\frac{M_{\rm mess}}{10^9\GEV}\right)^2 \left( \frac{1\GEV}{T_d} \right),
\end{eqnarray}
where $s_\text{after}$ and $s_\text{before}$ is the entropy density after or before the messenger decay, respectively, and $T_d$ is the decay temperature:
\begin{eqnarray}
T_d \simeq \left({\frac{45}{2\pi^2 g_*(T_d)}}\right)^{1/4}  \sqrt{M_P \Gamma_{\rm mess}} \sim 90 k_0 \left(\frac{61.75}{g_*(T_d)}\right)^{1/4} \left(\frac{M_\text{mess}}{10^9\GEV}\right)^{3/2}\GEV.
\end{eqnarray}
Thus, $D$ can be written as follows.
\begin{eqnarray}
D\simeq \frac{490}{90 k_0} \left(\frac{M_{\rm mess}}{10^9\GEV}\right)^{1/2} \left(\frac{61.75}{g_*(T_d)}\right)^{-1/4}
\end{eqnarray}
The gravitino abundance is the following:
\begin{eqnarray}
\Omega_{3/2} h^2 = (1/D)\, \Omega_{3/2}^{th} h^2 \simeq 0.1 \cdot \left(\frac{490}{D}\right) \left(\frac{m_{3/2}}{100 \KEV} \right) \left(\frac{230}{g_* (T_f)} \right).
\end{eqnarray}
Here, one can see that the observed relic abundance $\Omega_{3/2} h^2\simeq 0.12$ is naturally achieved 
without significant tuning.
It is emphasized that the abundance of the gravitino no longer depends on the reheating temperature
as long as the messengers were once in the thermal bath.
The sufficient baryon number can be produced thorough the leptogenesis for  $M_{N} \gtrsim 10^{11\mbox{-}12}$ GeV~\cite{baryon_lepto, Fujii:2002fv} and $T_R \gtrsim M_N$, where $M_N$ is the mass of the right-handed neutrino.

%%%%%%%%%%%%%%%
\begin{figure}[!t]
\begin{center}
\includegraphics[scale=1.1]{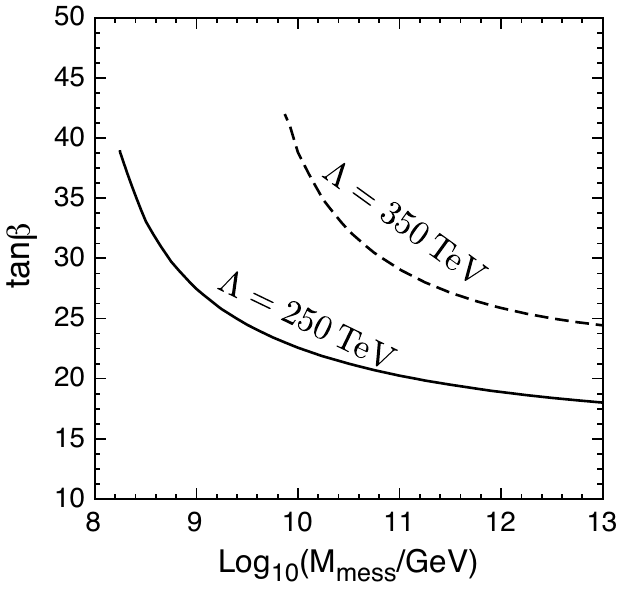}
\caption{
Predicted $\tan\beta$ for $B_\mu (M_{\rm mess})=0$, where the absence of the SUSY CP-problem is promised.
Here, $\mu<0$.
}
\label{fig:cpsafe}
\end{center}
\end{figure}
%%%%%%%%%

\section{Conclusion and discussion}

We have proposed a focus point gauge mediation model based on the product group unification.
It has been shown that the observed Higgs boson mass around 125 GeV 
is explained with a mild fine-tuning $\Delta \sim$\,70\,-\,130, an order of magnitude improvement over the minimal gauge mediation models. In addition,
the focus point naturally appears from in the PGU if the messenger field is an adjoint representation of $SU(5)$ gauge group;
unlike other focus point gauge mediation models, we do not need to choose the continuous parameters,
which control the focus point behavior. This enables our scenario to be very predictive. 
On the cosmological aspect, the gravitino can be dark matter 
without spoiling the success of the thermal leptogenesis. The observed baryon number of the universe is easily explained without the overproduction of the gravitino.

It should be noted that the SUSY CP-problem is solved in our framework 
with the vanishing Higgs $B$-term at the messenger scale.
In this case, all arguments of $A$-terms, gaugino masses and Higgs $B$-term are aligned to ${\rm arg}(\Lambda)$,\footnote{
The alignment is slightly deviated by corrections from gravity mediation of $\mathcal{O}(m_{3/2})$. 
This may give sizable contributions to electric dipole moments of leptons and quarks~\cite{gmsb_edm}.
}
 and hence, no dangerous CP-violating phases arise. Predicted values of $\tan\beta$ are around 20-40, depending on the messenger scale and the size of $\Lambda$, as shown in Fig.~\ref{fig:cpsafe}.

An unfortunate consequence of our model is that the masses of the SUSY particles are rather heavy, and it may be challenging to test it at the LHC,\footnote{
In our scenario, the next-to-lightest SUSY particle is stable at the collider time scale,
because of the gravitino mass larger than about 100 keV.
} 
even with the high luminosity running.
However, the relevant region with a mild fine-tuning is expected to be covered by the high energy upgrade of the LHC to 33\,TeV~\cite{33tev}.
The higgsino is always the next-lightest supersymmetric particle beyond the gravitino, 
and can be lighter than 500\,GeV. Therefore, it may be a target at the future $e^+ e^-$ linear collider experiments, such as ILC and CLIC.

Finally, let us comment on the proton decay, which can be a probe of our scenario. 
In the PGU, 
the colored Higgs multiplets are heavy without a difficulty, and hence, 
induced dimension five proton decay can be suppressed.
On the other hand, for the dimension six proton decay induced by $X$ and $Y$ gauge bosons, 
the decay rate of the proton can be ten times larger than that 
in the MSSM without any extra-particles~\cite{dim_6_proton_decay}. 
This is because we have the messenger field with a large representation of $SU(5)$, 
contributing the beta-functions of the standard model gauge couplings.
Therefore, the $SU(5)$ gauge coupling constant is larger than that of 
the standard SUSY GUT. 
Thus, the dimension six proton decay, in particular the $p \rightarrow e^+ \pi^0$ mode,  is a good target at the proposed Hyper-Kamiokande experiment \cite{Abe:2011ts}.

\section*{Acknowledgments}
Grants-in-Aid for Scientific Research from the Ministry of Education, Culture, Sports, Science, and Technology (MEXT), Japan, No. 26104009 (T. T. Y.); Grant-in-Aid No. 26287039 (T. T. Y.) from the Japan Society for the Promotion of Science (JSPS); and by the World Premier International Research Center Initiative (WPI), MEXT, Japan (H.M. and T. T. Y.).
H.\,M. is supported in part by the U.S. DOE under Contract DE-AC03-76SF00098, in part by the
NSF under grant PHY-1316783, in part by the JSPS Grant-in-Aid for
Scientific Research (C) (No.~26400241), Scientific Research on
Innovative Areas (No.~15H05887).
The research leading to these results has received funding
from the European Research Council under the European Unions Seventh
Framework Programme (FP/2007-2013) / ERC Grant Agreement n. 279972
``NPFlavour'' (N.\,Y.).
%%%%%%%%%%%%%%%

\appendix

\section{The soft SUSY breaking masses in gauge mediated SUSY breaking}

Here, we list the formulae for the gaugino and scalar masses in GMSB models. 
We consider two models: the one is the model with messenger multiplets, which are fundamental representations of $SU(2)_L$ and $SU(3)_c$, 
and the other is a model with messenger multiplet in the adjoint representation of $SU(5)$.

\subsection{Messenger in fundamental representation} \label{sec:mgmsb}

We consider a gauge mediation model with $N_D$ and $N_T$ pairs of the messenger multiplets transforming in the fundamental representation of $SU(2)_L$ and $SU(3)_c$. The relevant superpotential is given by
\begin{eqnarray}
W = (\lambda_D Z  + M_D ) \Psi_{D}^a \Psi_{\bar D}^a  +  (\lambda_T Z  + M_T ) \Psi_{T}^I \Psi_{\bar T}^I,
\end{eqnarray}
where $\Psi_D$ and $\Psi_T$ are $SU(2)_L$ doublet and $SU(3)_c$ triplet, respectively, and $U(1)_Y$ charges of $\Psi_D$ and $\Psi_T$ are taken as (-1/2) and (1/3). The index $a$ ($I$) runs $1$ to $N_D$ ($N_T$). 

Then, gaugino masses are given by
\begin{eqnarray}
M_{\tilde b} \simeq \frac{g_1^2}{16\pi^2} (\frac{3}{5} N_D + \frac{2}{5} N_T ) \Lambda, \ \ 
M_{\tilde w} \simeq \frac{g_2^2}{16\pi^2} N_D  \Lambda, \ \ 
M_{\tilde g} \simeq \frac{g_3^2}{16\pi^2} N_T  \Lambda,
\end{eqnarray}
where $\Lambda = \lambda_D F_Z/M_D = \lambda_T  F_Z/M_T$, provided that $\lambda_T=\lambda_D$ and 
$M_T=M_D$ hold at the GUT scale.
The SM gauge couplings of $SU(3)_c$, $SU(2)_L$ and $U(1)_Y$ are denoted by $g_3$, $g_2$ and $g_1$.
Here, $\lambda_{(D,T)} \left<Z\right> \ll M_{(D,T)}$ is assumed. Scalar masses are 
\begin{eqnarray}
m_{\tilde{Q}}^2 &\simeq& \frac{2}{(16\pi^2)^2}
\left[
\frac43 g_3^4 (N_T \Lambda^2  )
+ \frac34 g_2^4 (N_D \Lambda^2)
+ \frac35 g_1^4 (\tilde \Lambda_1^2) \frac{1}{6^2}
\right],
\nonumber \\
m_{\tilde{U}}^2 &\simeq& \frac{2}{(16\pi^2)^2}
\left[
\frac43 g_3^4 (N_T \Lambda^2   ) + \frac35 g_1^4 (\tilde \Lambda_1^2)
\left(\frac23\right)^2
\right],
\nonumber \\
m_{\tilde{D}}^2 &\simeq& \frac{2}{(16\pi^2)^2}
\left[\frac43 g_3^4 (N_T \Lambda^2)  + \frac35 g_1^4 (\tilde \Lambda_1^2) \frac{1}{3^2}
\right],
\nonumber \\
m_{\tilde{L}}^2 &\simeq& \frac{2}{(16\pi^2)^2}
\left[ \frac34 g_2^4 (N_D \Lambda^2) + \frac35 g_1^4 (\tilde \Lambda_1^2)
\frac{1}{2^2}
\right],
\nonumber \\
m_{\tilde{E}}^2 &\simeq& \frac{2}{(16\pi^2)^2}
\left[ \frac35 g_1^4 (\tilde \Lambda_1^2) \right],
\nonumber \\
m_{H_u}^2 &=& m_{H_d}^2 = m_{\tilde{L}}^2,
\end{eqnarray}
where $\tilde \Lambda_1^2 \equiv [(3/5) N_D+ (2/5) N_T] \Lambda^2$.

\subsection{Messenger in adjoint representation}

The messenger multiplet in the ${\bf 24}$ representation of $SU(5)$ is considered.
After $SU(5)$ is broken down to the SM gauge group, the superpotential in the messenger sector is written as
\begin{eqnarray}
W = (\lambda_8 Z  + M_8 ){\rm Tr}(\Sigma_8^2) + (\lambda_3 Z  + M_3 ){\rm Tr}(\Sigma_3^2) + (\lambda_X Z + M_X) \Psi_{X} \Psi_{\bar X},
\end{eqnarray}
where $\Sigma_8$ and  $\Sigma_3$ are adjoint representations of $SU(3)_c$ and $SU(2)_L$, respectively, and $\Psi_X$ ($\Psi_{\bar X}$) is the bi-fundamental (anti-bi-fundamental) representation of those SM gauge group with a $U(1)_Y$ charge of $-5/6$ $(5/6)$.
Then, gaugino masses are given by
\begin{eqnarray}
M_{\tilde b} \simeq \frac{g_1^2}{16\pi^2} (5 \Lambda_X), \ \ 
M_{\tilde w} \simeq \frac{g_2^2}{16\pi^2} (2 \Lambda_3 + 3 \Lambda_X), \ \ 
M_{\tilde g} \simeq \frac{g_3^2}{16\pi^2} (3 \Lambda_8 + 2 \Lambda_X),
\end{eqnarray}
where $\Lambda_8 = \lambda_8 F_Z/M_8$, $\Lambda_3 = \lambda_3 F_Z/M_3$ and $\Lambda_X=\lambda_X  F_Z/M_X$.  Scalar masses are 
\begin{eqnarray}
m_{\tilde{Q}}^2 &\simeq& \frac{2}{(16\pi^2)^2}
\left[
\frac43 g_3^4 (3\Lambda_8^2 + 2 \Lambda_X^2)
+ \frac34 g_2^4 (2\Lambda_3^2 + 3 \Lambda_X^2)
+ \frac35 g_1^4 (5 \Lambda_X^2) \frac{1}{6^2}
\right],
\nonumber \\
m_{\tilde{U}}^2 &\simeq& \frac{2}{(16\pi^2)^2}
\left[
\frac43 g_3^4 (3\Lambda_8^2 + 2 \Lambda_X^2) + \frac35 g_1^4 (5 \Lambda_X^2)
\left(\frac23\right)^2
\right],
\nonumber \\
m_{\tilde{D}}^2 &\simeq& \frac{2}{(16\pi^2)^2}
\left[\frac43 g_3^4 (3\Lambda_8^2 + 2 \Lambda_X^2)  + \frac35 g_1^4 (5 \Lambda_X^2) \frac{1}{3^2}
\right],
\nonumber \\
m_{\tilde{L}}^2 &\simeq& \frac{2}{(16\pi^2)^2}
\left[ \frac34 g_2^4 (2\Lambda_3^2 + 3 \Lambda_X^2) + \frac35 g_1^4 (5 \Lambda_X^2)
\frac{1}{2^2}
\right],
\nonumber \\
m_{\tilde{E}}^2 &\simeq& \frac{2}{(16\pi^2)^2}
\left[ \frac35 g_1^4 (5 \Lambda_X^2) \right],
\nonumber \\
m_{H_u}^2 &=& m_{H_d}^2 = m_{\tilde{L}}^2.
\end{eqnarray}

\section{Two-loop beta-functions for the gauge couplings}
Following Ref.~\cite{Martin:1993zk}, we show the two-loop beta-functions of the SM gauge couplings with $X$, $\bar X$, $\Sigma_3$ and $\Sigma_8$.
The contributions to the beta-functions from $\Sigma_3$ and $\Sigma_8$ are 
\begin{eqnarray}
\frac{d g_i}{d t} = \left(\frac{d g_i}{d t}\right)_{\rm MSSM} + \frac{\Delta b_i^{(1)}}{16\pi^2} g_i^3 + \frac{\Delta b_i^{(2)}}{(16\pi^2)^2} g_i^5,
\end{eqnarray}
where $(\Delta b_1^{(1)}, \Delta b_2^{(1)}, \Delta b_3^{(1)})=(0, 2, 3)$ and $(\Delta b_1^{(2)}, \Delta b_2^{(2)}, \Delta b_3^{(2)})=(0, 24, 54)$.
The contributions from $X$ and $\bar X$ are 
\begin{eqnarray}
\frac{d g_i}{d t} = \left(\frac{d g_i}{d t}\right)_{\rm MSSM} + \frac{\Delta' b_i^{(1)}}{16\pi^2} g_i^3 + \frac{\Delta' b_{ij}^{(2)}}{(16\pi^2)^2} g_i^3 g_j^2,
\end{eqnarray}
where $(\Delta' b_1^{(1)}, \Delta' b_2^{(1)}, \Delta' b_3^{(1)})=(5, 3, 2)$ and
\begin{eqnarray}
\Delta' b_{ij}^{(2)} = 
\left(
\begin{array}{ccc}
 \frac{25}{3} & 15   &\frac{80}{3}   \\
5  & 21  &   16\\
\frac{10}{3}  & 6  &   \frac{68}{3}
\end{array}
\right).
\end{eqnarray}

\end{document}